# Social Signals in the Ethereum Trading Network

Shahar Somin[1,3], Goren Gordon[2,3], and Yaniv Altshuler[1,3]

[1] MIT Media Lab, MA, USA
{shaharso, yanival}@media.mit.edu
[2] Curiosity Lab, Industrial Engineering Department, Tel Aviv University, Israel
goren@gorengordon.com
[3] Endor Ltd.

**Abstract.** Blockchain technology, which has been known by mostly small technological circles up until recently, is bursting throughout the globe, with a potential economic and social impact that could fundamentally alter traditional financial and social structures. Issuing cryptocurrencies on top of the Blockchain system by startups and private sector companies is becoming a ubiquitous phenomenon, inducing the trading of these crypto-coins among their holders using dedicated exchanges.
Apart from being a trading ledger for tokens, Blockchain can also be observed as a social network. Analyzing and modeling the dynamics of the "social signals" of this network can contribute to our understanding of this ecosystem and the forces acting within in.
This work is the first analysis of the network properties of the ERC20 protocol compliant crypto-coins' trading data. Considering all trading wallets as a network's nodes, and constructing its edges using buy–sell trades, we can analyze the network properties of the ERC20 network. Examining several periods of time, and several data aggregation variants, we demonstrate that the network displays strong power-law properties. These results coincide with current network theory expectations, however nonetheless, are the first scientific validation of it, for the ERC20 trading data.
The data we examined is composed of over 30 million ERC20 tokens trades, performed by over 6.8 million unique wallets, lapsing over a two years period between February 2016 and February 2018.

## 1 Introduction

The Ethereum Blockchain, launched in July 2015 [1], is a public ledger that keeps records of all Ethereum related transactions. It is shared between all participants and is based on a reward mechanism as an incentive for users to run the transactions network. A key characteristic of the Blockchain network is its heavy reliance on cryptography to secure the transactions, addressed as the consensus mechanism. Each account consists of a public and private key duo, where the private key is used to digitally sign each account's transactions, and the public key can be used by all Blockchain participants in order to verify the transaction's validity, in a rapid, decentralized and transparent way.

The ability of the Ethereum Blockchain to store not only ownership, similarly to Bitcoin, but also execution code, in the form of *"Smart Contracts"*, has recently led to the creation of a large number of new types of "tokens", based on the Ethereum ERC20 protocol. These tokens are "minted" by a variety of players, for a variety of reasons, having all of their transactions carried out by their corresponding Smart Contracts, publicly accessible on the Ethereum Blockchain.

In this regards, the Ethereum Blockchain's transactions, and ERC20 transactions in particular, constitute a decentralized record of interactions among participants, with two interesting properties that distinguish it from most of the traditional interaction collections (such as social network activities, phone-call records, financial bank transactions):

- **Unlimited number of wallets** — The Ethereum private key mechanism enables any participant to create an unlimited amount of unique "wallets". Whereas the participant can control all of these wallets easily, it is impossible for an outside observer to explicitly associate the wallets to each other (with the exception of an implicit association, through a careful data analysis work, as can be seen in [2]). This can be compared to a mobile phone network, in which every participant may hold an infinite amount of different identities, addressed as phone numbers, all of which can be used at will. Had this property existed in reality, it would likely render most of recent seminal works in this field (such as [3–8] and many more) highly impractical, if not entirely obsolete, as demonstrated in [9].
- **Unlimited number of tokens** — The ability of participants to create not only new wallet addresses, but also an unlimited number of new *tokens* turns the Ethereum network from a single faceted means of communication of storage and execution related transactions, to a multi faceted (and in fact, an infinitely faceted) one, comprised of many different types of interactions, whose nature widely varies from payment, through decentralized trading in GPU resources [10], and to consumption of behavioral predictions [12].

As a result, the ERC20 ecosystem and the multitude of transactions it consists of, constitutes one of the most fascinating examples for decentralized networks. However, to this day there has not been any in-depth analysis of the ERC20 tokens network properties .

This work is the first attempt to analyze the ERC20 tokens through a network theory prism. We study two years of ERC20 transactions over the Ethereum Blockchain, by forming a social network from the participants and their corresponding monetary actions. We show that the ERC20 tokens data, despite being infinitely faceted and potentially comprised of unlimited amount of single-serving wallet addresses, still strongly displays several key properties known in network theory research to characterize sets of human interactions. The direct potential implication of our discovery is that the ERC20 tokens data is likely to therefore also comply with additional known network properties – leading the way for the development of an abundance of predictive and descriptive techniques for the ERC20 tokens transactions, based on known network theory oriented approaches from other domains.

The rest of the paper is organized as follows: Section 2 contains background on the topics of this work and review of previous literature related to it. In Section 3 we thoroughly describe the methodology that was used in this work, whereas the results are discussed in Section 4. Concluding remarks and discussion regarding future work appear in Section 5.

## 2 Background and Related Work

Blockchain's ability to process transactions in a trust-less environment, apart from trading its official cryptocurrency, the *Ether*, presents the most prominent framework for the execution of "*Smart Contracts*" [13]. Smart Contracts are computer programs, formalizing digital agreements, automatically enforced to execute any predefined conditions using the consensus mechanism of the Blockchain, without relying on a trusted authority. They empower developers to create diverse applications in a Turing Complete Programming Language, executed on the decentralized Blockchain platform, enabling the execution of any contractual agreement and enforcing its performance.

Moreover, Smart Contracts allow companies or entrepreneurs to create their own proprietary tokens on top of the Blockchain protocol [14]. These tokens are often pre-mined and sold to the public through Initial Coin Offerings (ICO) in exchange of Ether, other crypto-currencies, or *Fiat Money*. The issuance and auctioning of dedicated tokens assist the venture to crowd-fund their project's development, and in return, the ICO tokens grant contributors with a redeemable for products or services the issuer commits to supply thereafter, as well as the opportunity to gain from their possible value increase due to the project's success.

The most widely used token standard is Ethereums *ERC20* (representing Ethereum Request for Comment), issued in 2015. The protocol defines technical specifications giving developers the ability to program how new tokens will function within the Ethereum ecosystem. In order to comply with the ERC20 standard, the token must adhere to rules regarding the form it will be transferred between addresses and the manner in which data within the token is accessed. The Contract stores the addresses of its corresponding token owners, alongside with the amount of owned tokens, and allows token transfers only if the sender proves ownership of the private key associated to the Contract address.

This brand new market of ERC20 compliant tokens is fundamental to analyze, as it is becoming increasingly relevant to the financial world. Issuing tokens on top of the Blockchain system by startups and other private sector companies is becoming a ubiquitous phenomenon, inducing the trade of these crypto-coins to an exponential degree. Since 2017, Blockchain startups have raised over 7 Billion dollars through ICOs. Among the largest offerings, *Tezos* raised $232M for developing a smart contracts and decentralized governance platform; *File-*

*coin* raised $205M to deploy a decentralized file storage network; *EOS* raised over $185M to fund scalable smart contracts platform, *Bancor* managed to raise $153M for deploying a Blockchain-based prediction market, *Kin* $98M to build a decentralized social network and communication platform and *Blockstack* raised $52M towards a decentralized browser, identity and application ecosystem.

Apart from being formed by countless stake-holders and numerous tokens, the ERC20 transactional data also presents full data of prices, volumes and holders distribution. This, alongside with daily transactions of anonymised individuals is otherwise scarce and hard to obtain due to confidentiality and privacy control, hence providing a rare opportunity to analyze and model financial behavior in an evolving market over a long period of time.

In the past two decades, network science has exceedingly contributed to multiple and diverse scientific disciplines. Applying network analysis and graph theory have assisted in revealing the structure and dynamics of complex systems by representing them as networks, including social networks [15–17], computer communication networks [18], biological systems [19], transportation [20, 21], IOT [22], emergency detection [23] and financial trading systems [24–26].

Most of the research conducted in the Blockchain world, was concentrated in Bitcoin, spreading from theoretical foundations [27], security and fraud [28, 29] to some comprehensive research in network analysis [30–32]. The world of Smart contracts has recently inspired research in aspects of design patterns, applications and security [33–36], policy towards ICOs has also been studied [14]. However, the comprehensive analysis of ERC20 tokens, with emphasis on the investigation of the transaction graph built from their related activity on the Blockchain, is still lacking.

In this paper we aim to examine how this prominent field can enhance the understanding of the underlying structure of the ERC20 tokens trading data.

## 3 Methodology

### 3.1 Data

In order to preserve anonymity in the Ethereum Blockchain, personal information is omitted from all transactions. A User, represented by their wallet, can participate in the economy system through an address, which is attained by applying *Keccak-256* hash function on his public key. The Ethereum Blockchain enables users to send transactions in order to either send Ether to other wallets, create new Smart Contracts or invoke any of their functions. Since Smart Contracts are scripts residing on the Blockchain as well, they are also assigned a unique address. A Smart Contract is called by sending a transaction to its address, which triggers its independent and automatic execution, in a prescribed manner on every node in the network, according to the *data* that was included in the triggering transaction.

Smart Contracts representing ERC20 tokens comply with a protocol defining the manner in which the token is transferred between wallets. Among these requirements, is the demand to implement a *transfer* method, which will be used for transferring the relevant token from one wallet to another. Therefore, each transfer of an ERC20 token will be manifested by a wallet sending a transaction to the relevant Smart Contract. The transaction will encompass a call to the *transfer* method in its *data* section, containing the amount being transferred and its recipient wallet. Each such token transfer results in altering the 'token's balance', which is kept and updated in its corresponding Smart Contract's storage.

We obtain the ERC20 transactions basing on the further requirement of the ERC20 protocol, demanding that each call to the *transfer* method will be followed by sending a *Transfer* event and updating the event's logs with all relevant information regarding the token transfer. We therefore call an Ethereum full node's JSON API and fetch all logs matching to the *Transfer* event structure [37]. Parsing these logs result in the following fields per transaction: *Contract Address* - standing for the address of the Smart Contract defining the transferred token, *Value* - specifying the amount of the token being transferred, *Sender* and *Receiver* addresses, being the wallet addresses of the token's seller and buyer, correspondingly.

We have retrieved all ERC20 tokens transactions spreading between February 2016 and February 2018, resulting in $30,347,248$ transactions and $18,517$ token address. Due to the restriction on changing and tempering Smart Contracts, any modification made to a token's designated Smart Contract involves a definite change in it's associated Contract Address. As a result, a token can change addresses throughout it's lifespan, though for any point in time, it will only be assigned to a single relevant *Contract Address*. Therefore, the above mentioned amount of unique contract addresses serves merely as an upper bound to the amount of unique tokens. Since we do not restrict ourselves to a specific type of token, but observe the network as a whole trading system, this non-unique identification of tokens doesn't affect our analysis of the network.

The dataset of ERC20 tokens transactions is extremely diverse and wide-ranging, where not only any ERC20 token might correspond to multiple contract addresses, and therefore being considered as various different tokens by our analysis, but also the characteristics of the different tokens are extremely varied. For instance, the tokens differ in their age, their economic value, activity volume and number of token holders, some merely serve as test-runs, others aren't tradable in exchanges yet, and some, according to popular literature, are frauds, all residing next to actual real-world valuable tokens.

### 3.2  Graph Analysis

In order to perceive the network's structure and assess the connectivity of its nodes, one should examine the network's degree distribution, considering both in-degree and out-degree, indicating the number of incoming and outgoing connections, correspondingly. The degree distribution $P(k)$ signifies the probability that a randomly selected node has precisely the degree $k$.

In random networks of the type studied by Erdös and Rényi [38], where each edge is present or absent with equal probability, the nodes' degrees follow a *Poisson* distribution. The degree obtained by most nodes is approximately the average degree $\bar{k}$ of the network. These properties are also manifested in dynamic networks [39]. In contrast to random networks, the nodes' degrees of social networks (such as the Internet or citation networks) often follow a *power law* distribution [40]:

$$P(k) = k^{-\alpha} \tag{1}$$

The power law degree distribution indicates that there is a non-negligible number of extremely connected nodes even though the majority of nodes have small number of connections. Therefore the degree distribution has a long right tail of values that are far above the average degree. Power law distributions can be found in many real networks, Newman [17] summarized several of them, including word frequency, citations, telephone calls, web hits, or the wealth of the richest people.

## 4  Results

As discussed, we study an extremely diverse and wide-ranging dataset. In order to present a first glimpse on the diversity of ERC20 tokens transactional data, we explore the distribution of token popularity, in terms of buyers and sellers amount. As Figure 1 reflects, ERC20 tokens' popularity follows a power-law distribution, thereby expressing the diversity of token holders along a 2 years period, between February 2016 and February 2018. Particularly, it can be seen that most tokens are traded by an extremely small amount of users and on the other hand, a few popular tokens exist, traded by a very large group of users during the examined timespan.

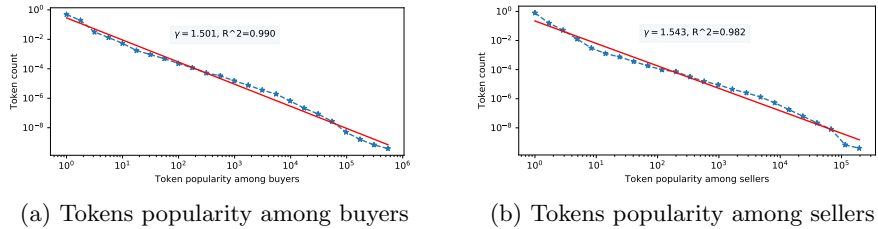

(a) Tokens popularity among buyers    (b) Tokens popularity among sellers

Fig. 1: Crypto-coins popularity over a two year period lapsing from February 2016 to February 2018. Depicting the probability a coin would have certain amount of buyers (left panel) and sellers (right panel), both demonstrating a power-law distribution.

We further aim to examine whether the ERC20 network satisfies the known characteristics of other real-world networks, first and foremost examining its degree distribution. We therefore construct the following directed graph, $G_{FT}(V, E)$, standing for *ERC20 Full Transactions Graph*, including all transactions in the timespan between February 2016 and February 2018. The resulting graph consists of $6,890,237$ vertices and $17,392,610$ edges.

The set of vertices $V$ consists of all ERC20 trading wallets in this period, where any vertex $u$ represents a trading wallet $w_u$. Out-going edges depict transactions in which wallet $w_u$ sold any type of ERC20 token to other wallets, and in-coming edges to $u$ are formed as result of transactions in which $w_u$ bought any ERC20 token from others. Formally, $E \subseteq V \times V$ s.t.:

$$E := \{(u,v) \| \text{ wallet } w_u \text{ sold any ERC20 token to } w_v\} \quad (2)$$

Out-degree of vertex $u$ represents the number of unique wallets buying tokens from $w_u$ and its in-degree depicts the number of unique wallets selling tokens to it.

Surprisingly, despite the great variance between the traded tokens in the network, we discovered that the degree distribution depicts a strong power-law pattern, as presented in Figure 2. Hence the *ERC20 Full Transactions Graph*, $G_{FT}$, displays similar connectedness structure to other real-world networks, presenting a non-negligible number of extremely connected nodes even though the majority of nodes have small number of connections, both in buying and selling transactions.

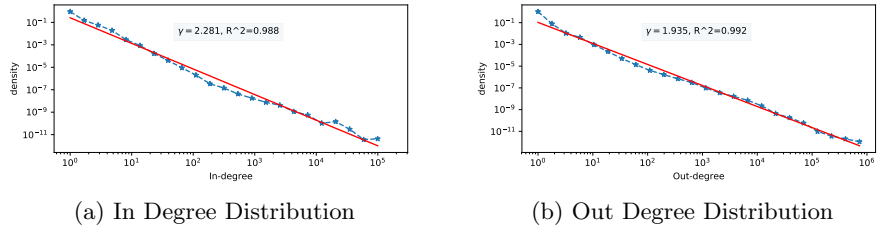

(a) In Degree Distribution    (b) Out Degree Distribution

Fig. 2: Analysis of Blockchain network dynamics for a 2 years period from February 2016 to February 2018. The networks nodes represent ERC20 wallets and edges are formed by ERC20 buy-sell transactions. Outgoing degree of a node reflects the number of unique wallets receiving funds from that node, and vice-versa for incoming degree. Both outgoing and incoming degrees present a power-law distribution.

We have additionally analyzed ERC20 transaction graphs based on varying length periods between 3 days to 3 months, and validated our findings across 20 different points in time. We have observed that in all cases the power-law degree distribution is preserved and presents roughly similar $\gamma$ values. We omit these results from the current version, due to space limitations, and they will appear in a future, extended version.

## 5   Concluding Remarks and Future Work

In this paper, we have demonstrated for the first time that the ERC20 tokens transactional data displays several properties known to be associated with networks that are comprised of human interactions, and social networks specifically. This occurs despite the fact that the Blockchain protocol enables the creation of an unlimited number of "tokens", causing diverse sub-domains to reside together over the same protocol, and regardless of an unlimited amount of wallets, resulting in different identities controlled by a single individual.

Specifically, we have modeled the transactions as a network that is comprised of wallets, connected through transactions, and found that the degree distribution of nodes in the network presents a power-law pattern. In addition, we have shown that tokens popularity among buyers and sellers also follows a power-law model. These preliminary results indicate that (somewhat surprisingly) despite its diversity, ERC20 data presents a social behavior. This leads us to explore whether other aspects of network theory can emerge from this data. Such fields include short path lengths and clustering coefficient analysis [41], centrality measures [42], connected components behavior [5] and community structure study [43]. We have already been able to demonstrate some of these phenomena using the ERC20 data as well, however they were not included in this work due to space considerations.